\documentclass{aa} 

\usepackage{graphicx}
\usepackage{txfonts}
\usepackage[colorlinks=true,urlcolor=blue,citecolor=blue,linkcolor=blue,bookmarks=true]{hyperref}

\newcommand{\BH}{{\rm BH}}
\newcommand{\g}{{\rm g}}

\begin{document}

\title{Tidal disruption and evaporation of rubble-pile and monolithic bodies as a source of flaring activity in Sgr A$^\star$}

    \authorrunning{Zhou et al}

    \author{Wen-Han Zhou\thanks{Present address: Department of Earth and Planetary Science, The University of Tokyo, Tokyo, Japan}
          \inst{1,2}
          \and
          Yun Zhang
          \inst{3}
          \and
          Jiamu Huang
          \inst{4}
          \and
          Douglas N.C. Lin
          \inst{5,6}
          }

   \institute{
   Universit\'e C\^ote d'Azur, Observatoire de la C\^ote d'Azur, CNRS, Laboratoire Lagrange, Nice, France
   \\
   \email{wenhan.zhou@oca.eu}
   \and
   Department of Earth and Planetary Science, The University of Tokyo, Tokyo, Japan
    \and
    Department of Climate and Space Sciences and Engineering, University of Michigan, Ann Arbor, MI 48109, USA
    \and
    Department of Physics, University of California, Santa Barbara, USA
    \and
    Department of Astronomy and Astrophysics, University of California, Santa Cruz, USA
    \and
    Institute for Advanced Studies, Tsinghua University, Beijing, China}

  \abstract
    {Sgr A$^\star$, the supermassive black hole at the center of the Milky Way, exhibits frequent short-duration flares (e.g., with luminosity $> 10^{34}~\rm erg~s^{-1}$) across multiple wavelengths. The origin of the flares is still unknown.}
    {We revisited the role of small planetary bodies, originally from the stellar disk, and their tidally disrupted fragments as a source of flaring activity in Sgr A$^\star$. In particular, we refined previous models by incorporating material strength constraints on the tidal disruption limit and by evaluating the evaporation dynamics of the resulting fragments.}
    {We analyzed the tidal fragmentation and gas-induced fragmentation of small planetary bodies with rubble-pile and monolithic structures. Using constraints from recent space missions (e.g., NASA's OSIRIS-REx and JAXA's Hayabusa2 missions), we estimated the survivability of fragments under aerodynamic heating and computed their expected luminosity from ablation, modeled as fireball flares analogous to meteor events.}
    {We find that planetary fragments can approach as close as $8 \, R_\bullet$ due to material strength, where $R_\bullet$ denotes the gravitational radius consistent with flare locations inferred from observations. The fireball model yields luminosities from $10^{34 }$ to $10^{36}$~erg/s for fragments whose parent bodies are a few kilometers in size. The derived flare frequency--luminosity distribution follows a power law with the power index 1.83, in agreement with observed values (1.65--1.9), while the flare duration scales as $t_{\rm f} \propto L^{-1/3}$, consistent with observational constraints. We considered the discovered young stars around Sgr A$^\star$ as the planetary reservoir. Given a small-body population analogous in mass to the primordial Kuiper belt and the common existence of close-in super-Earths as well as long-period Neptunes, we show that this planetary reservoir can provide an 
    adequate supply for the observed flares.
    }
    {The tidal disruption and thermal evaporation of small bodies offer a plausible explanation for the observed flare properties of Sgr A$^\star$. }
    

   \maketitle
%
\section{Introduction}

Sgr A$^\star$ is a supermassive black hole (SMBH) with 
 mass $M_\bullet = 4 \times 10^6 M_\odot$ \citep{Genzel1997, Ghez1998} and 
 gravitational radius $R_\bullet = G M_\bullet /c^2=6 \times 10^{11}$ cm.  It
exhibits variability across multiple wavelengths, including the near-infrared (NIR) \citep{Genzel2003, YZadeh2025}, mid-infrared (MIR) \citep{vFellenberg2025}, submillimeter \citep{Zhao2003}, and X-ray \citep{Baganoff2001, Nowak2012, Ponti2017}. Over the past two decades, observations of Sgr A$^\star$ have revealed that it undergoes frequent short-lived flares every few hours, during which its X-ray luminosity rises to $10^{35}$–$10^{36},\mathrm{erg~s^{-1}}$, its NIR emission peaks at $\sim10^{35}~\mathrm{erg~s^{-1}}$ \citep{Genzel2003,YZadeh2025}, its MIR output reaches $\sim10^{34}~\mathrm{erg~s^{-1}}$ \citep{vFellenberg2025}, and other bands can show comparable enhancements of order $10^{35}~\mathrm{erg~s^{-1}}$, all of which lie one to two orders of magnitude above the quiescent level (typically $<10^{33} ~\mathrm{erg~s^{-1}}$). The origin of these flares remains an open question; the   proposed mechanisms include magnetic reconnection within the accretion flow \citep[e.g.,][]{Markoff2001, DEden2010} and sudden increases in accretion \citep[e.g.,][]{Liu2002}. Another proposed explanation involves the tidal disruption of small bodies (e.g., asteroids and comets) near Sgr A$^\star$ \citep{Cadez2008, Kostic2009, zubovas2012}.

A population of small bodies, such as asteroids, comets, and planetesimals, is likely to exist in the Galactic center, originating from the dense stellar environment and fragmented planetary systems \citep{Nayakshin2012}. These objects may have formed in protoplanetary disks similar to that
suggested for the G-clouds \citep{owen2023}.
When these objects enter the tidal sphere of Sgr A$^\star$, they experience extreme gravitational stresses that lead to partial or complete disintegration. The resulting debris, which has a significantly larger surface area than the original bodies, interacts with the hot accretion flow and gets evaporated, producing transient emissions via synchrotron radiation by accelerated electrons from asteroid materials \citep{zubovas2012}. The timescales of these flares, ranging from tens of minutes to a few hours, are consistent with the expected evaporation timescales of disrupted asteroidal debris.

For this work we revisited the hypothesis that the tidal disruption of small bodies contributes to the observed flaring activity of Sgr A$^\star$. Here, we define small bodies as minor planets (i.e., asteroids, comets, and planetesimals) and planetary fragments that break apart near the Roche limit. Previous studies  oversimplified the modeling of small-body disruption by either neglecting the rubble-pile structure of asteroids \citep{Kostic2009} or ignoring material strength \citep{zubovas2012}, leading to an overestimation of the tidal disruption distance. We refined these estimates by incorporating constraints on asteroid material strength derived from recent space missions, such as JAXA’s Hayabusa series and NASA’s OSIRIS-REx mission. Additionally, we consider the role of gas pressure in the disintegration process. Recent JWST NIR observations have reported subflares with continuously varying brightness, followed by a burst of flaring activity \citep{YZadeh2025}. We suggest that this behavior may be analogous to fireball flares, where evaporating meteor fragments produce transient emissions. Furthermore, we estimate the delivery rate of small bodies into the vicinity of Sgr A$^\star$ by considering their detachment from stars on eccentric orbits.

The paper is organized as follows. In Sect.~\ref{sec2} we describe the fragmentation of planetary bodies due to tidal forces and gas pressure. In Sect.~\ref{sec3} we analyze the evaporation processes of these objects and examine the resulting luminosity and spatial distribution of flares. In Sect.~\ref{sec4} we evaluate the existence of small bodies in the stellar disk and estimate the occurrence rate of their tidal disruption based on the size required to produce the observed flare energy. In Sect.~\ref{sec5} we summarize our findings.

\section{Breakup of planetary bodies}
\label{sec2}

In the vicinity of a black hole, due to resonant relaxation, some young stars very close to Sgr A$^\star$ and old stars in the nuclear cluster endure eccentricity excitation and undergo tidal disruption when they venture very close to the  supermassive black hole.  Along the way, they release planets and small bodies around them.  This supply of freely floating  planets compensates the loss of low-mass objects due to mass segregation.  These detached planets and small bodies lose their gaseous envelope when they venture inside the conventional tidal disruption radius and downsize when the material strength of their residual solid cores can no longer withstand the tidal tearing by the supermassive black hole or the gas pressure. Thus, tidal forces and gas pressure impose an upper limit on the maximum size of small bodies that can survive in this environment.

We assume that objects smaller than 100 m in radius are monolithic bodies, based on the fast spin rates observed for similarly-sized objects in the Solar System. Objects between 100 and 10~km are considered rubble-pile bodies, which are gravitational aggregates composed of smaller monolithic fragments (each with $r < 100~$m) bound together by self-gravity and material cohesive strength \citep{Walsh2018}. As we  demonstrate here,
thanks to material cohesion, rubble-pile objects retain partial structural integrity and continue to downsize after they cross the traditional Roche limit ($\sim$$30 R_{\bullet}$). Due to the combined effects of tidal disruption and gas pressure overcoming material cohesion, their closest approach to Sgr A$^\star$ is approximately 22$R_\bullet$, within which they are reduced to smaller monolithic bodies. These monolithic objects possess sufficiently high material strength to withstand the tidal forces and gas pressures in the vicinity of the black hole.

\subsection{Tidal downsizing}
\label{sec:tidal_disruption}

\begin{figure}
    \centering
    \includegraphics[width=0.5\textwidth]{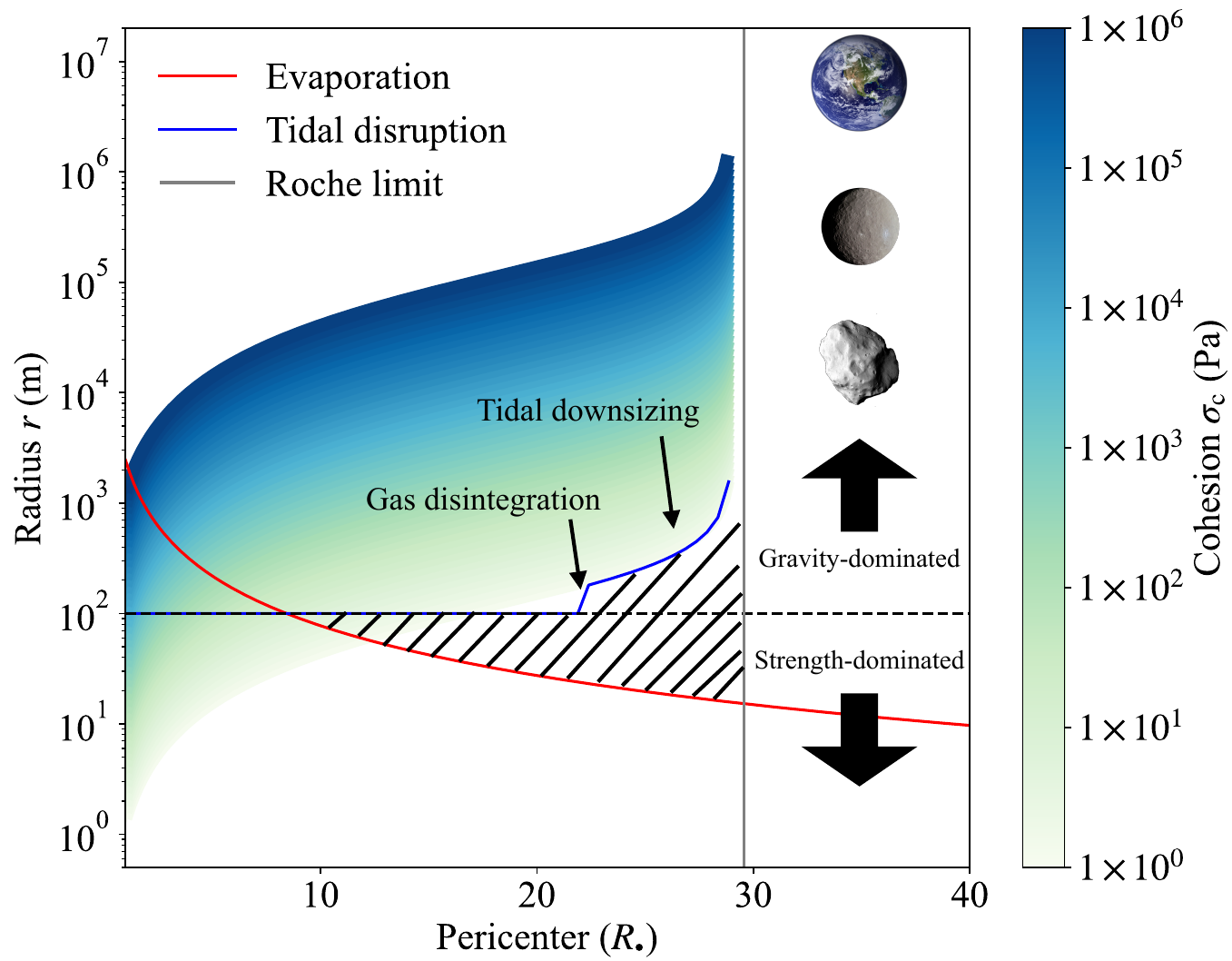}
    \caption{Tidal downsizing and evaporation of planetary bodies. The maximum size of a planetary body surviving tidal forces is denoted by the gradient-filled regions;  the color denotes the cohesive strength. The blue curve traces the maximum size of survivable small bodies under the tidal effect. Rubble piles ($>$100~m) are typically assumed to have a cohesion of 1~Pa, whereas monolithic bodies ($<$100~m) exhibit much higher cohesion, often exceeding $10^4~$Pa. The vertical gray line marks the classical Roche limit. The red line represents the minimum size of survivable small bodies under evaporation via gas drag during a single flyby. The shaded region represents the parameter space where small bodies can persist against both tidal disruption and hydrodynamic ablation. }
    \label{fig:asteroid_survival}
\end{figure}

A planetary object undergoes tidal disruption once it crosses the tidal disruption radius. For a gravity-dominated object, the tidal disruption radius is \citep{Sridhar92}
\begin{equation}
    R_{\rm t,grav} = 1.05 \left( \frac{M_\bullet}{\rho} \right)^{1/3} = 30 \left(\frac{\rho}{1.5\rm~g/cm^3} \right)^{-1/3} R_{\bullet}.
\label{eq:rgrav}
\end{equation}
Here $\rho$ is the bulk density of the object.
For a strength-dominated object, the tidal disruption radius is \citep{Holsapple07,Holsapple08, Zhang20}
\begin{equation}
    R_{\rm t,str} = \left( {\frac{\sqrt{3}}{4\pi}} \right)^{1/3} \left( \frac{5k}{4\pi G r^2 \rho^2} + s \right)^{-1/3} \left( \frac{M_\star}{\rho}\right)^{1/3},
\label{eq:rstr}
\end{equation}
where 
\begin{align}
    & k = \frac{6 \sigma_{\rm c} \cos \phi}{\sqrt{3} (3-\sin \phi)}  \\
    & s = \frac{2 \sin \phi}{\sqrt{3}(3-\sin \phi)}.
\end{align}
Here $r$ is the radius of the planetary object. The cohesive strength, $\sigma_{\rm c}$, can range from 0.1~Pa (e.g., between particles in rubble piles) to 10~MPa (e.g., monolithic rocks), and the frictional angle $\phi$ is typically in the range of $25^\circ$ to $45^\circ$ \citep{Zhang20}. {{By definition, $\tan \phi$ is the ratio of shear to normal stress at which the material begins to fail (e.g., slide, flow, or collapse), representing the internal resistance of granular material to shear deformation.}} The maximum radius of tidal fragments is a decreasing function of the orbital radius $R$:
\begin{equation}
    r_{\rm max} = \left( \frac{5k}{4\pi G \rho^2 ({\sqrt{3}M_\BH}/{4\pi \rho R^3} - s )} \right)^{1/2}.
\end{equation}
This is  called the tidal downsizing process, which generates numerous small bodies.
Figure~\ref{fig:asteroid_survival} shows the maximum size a small body can survive as a function of the pericenter distance, for different cohesion strengths $\sigma_{\rm c}$.

By analogy with small Solar System bodies, the objects orbiting the SMBH are also rubble piles; they exhibit cohesive strengths of merely a few pascals. According to \citet{Sanchez2014}, the cohesive strength of Solar System rubble-pile small bodies is 
\begin{equation}
    \sigma_{\rm c,rp} \sim 3 \times 10^{-4} \left( {1 {\rm ~m} \over r_{\rm p}} \right)\rm Pa,
\end{equation}
where $r_{\rm p}$ is the particle size. For micron-sized surface particles, the cohesion can reach a few hundred pascals. Since the rubble piles ($r_{\rm p}>100~$m) are made up of grains of a broad size range (e.g., from microns to meters or larger), estimates of real rubble-pile asteroids suggest cohesive strengths ranging from a few pascals to hundreds of pascals \citep{Rozitis2014,Scheeres10,Zhang18}. A structural analysis of Bennu, the target of NASA's OSIRIS-REx mission, implies near-zero surface cohesion, as inferred from the forces measured during surface sampling operations \citep{Walsh2022}. Additionally, the low-speed mobilization of ejecta deposited near the crater suggests a cohesion of less than $2~$Pa \citep{Perry2022}. The structure of the crater produced in the impact experiment on the asteroid Ryugu, the target of JAXA's Hayabusa2, suggests $\sigma_c \geq 0.05$~Pa \citep{Jutzi2022}.  The low  $\sigma_c$ values of Bennu and Ryugu could be the result of their  carbon-dominated compositions as both are classified as C-type asteroids based on the spectra. The S-type (silicate-dominated) asteroids may exhibit higher cohesive strengths. For example, the asteroid Didymos, the target of NASA's DART mission, is estimated to have a cohesive strength $\sim 20~$Pa \citep{Zhang2021b, Agrusa2022b}. Since C-type asteroids are abundant and typically located closer to comets than S-type asteroids, we assume $\sigma_{\rm c,rp} \sim 1 ~\rm Pa$ for rubble-pile small bodies. As shown in Fig.~\ref{fig:asteroid_survival}, the large rubble piles ($>100~$m) begin to fragment when crossing $R_{\rm Roche} \sim 30~R_{\bullet}$. This downsizing process continues until their size is reduced to $~\sim 100~$m, where they become monolithic bodies with much higher material strength. The closest distance for rubble piles to survive is $\sim$$ 18~R_\bullet$.

For monolithic bodies, the dependence of material strengths on size is still uncertain. Based on the observed meteoroids, the material strength can be described by 
\begin{equation}
\label{eq:sigma_c_mn}
    \sigma_{\rm c,mn} \sim \sigma_0 \left( r \over r_0 \right)^{-3\alpha}.
\end{equation}
Here $\alpha$ could vary from 0.05 to 0.5 \citep{Popova2011}. We assume $\alpha \sim 0.2$. An analysis of the size and depth of craters observed on boulders on the asteroid (101955) Bennu suggests $\sigma_{\rm c,mn} \sim 0.44-1.70~$MPa for meter-sized boulders \citep{Ballouz2020}. The boulders on the asteroid Ryugu suggest a lower strength 0.2-0.28~MPa \citep{Grott2019}. Constraints by the spin limit of asteroids \citep{Zhang2021} give $\sigma_0 = 10~$kPa, $r_0 = 0.5~$km, and $\alpha = 1/6$.
Despite the considerable uncertainty associated with material strength, various estimation methods consistently indicate that the material strength is higher than $>10~$kPa for small bodies with radii smaller than 100~m. Therefore, these monolithic objects can reach the vicinity of the black hole (e.g., $\ll 18~R_\bullet$).

\subsection{Disintegration caused by gas pressure}
\label{sec:gas_pressure}

However, the gas pressure can disintegrate the object if the gas pressure exceeds the object's bounding strength. Assuming disk accretion, the gas density $\rho_{\rm g}$ at a distance $R$ is 
\begin{equation}
\begin{aligned}
    \rho_{\rm g} &\simeq \frac{\dot M R}{2 \pi H^3 \alpha_{\rm d} v_{\rm ff}} \\
    &= 8 \times 10^{-14} \left({\dot M \over 10^{-8} {M_\odot ~\rm yr^{-1}}} \right)
    \left({R^3 \over \alpha_{\rm d} H^3} \right) \left({R \over R_\bullet} \right)^{-3/2}  \rm kg~m^{-3}.
\end{aligned}
\end{equation} 
Here $v_{\rm ff} = \sqrt{G M/R}$ is the free-fall velocity. The scaled height of the disk near the black hole, $H/R$, can be approximated as 1  \citep{Narayan1998}. The viscosity parameter $\alpha_{\rm d} $ generally remains constant for magnetically arrested disks, while it decreases with  distance for the weakly magnetized flow \citep{Liska2020, Chatterjee2022}. A recent general relativistic magnetohydrodynamic simulation on Sgr A$^\star$ suggests a constant $\alpha_{\rm d}$ \citep{Chatterjee2023}. Here we adopted $\alpha_{\rm d} \sim 0.3$, as suggested by \citet{Narayan1998}. The parameter $\dot M$ is the accretion rate. Observations of Faraday rotation in polarized millimeter emissions suggest a range between  $10^{-7}$ and $10^{-9}$ solar masses per year \citep{Bower2003, Marrone2006}. Further support comes from the Event Horizon Telescope (EHT) \citep{EHT2023}. Modeling suggests that the accretion rate near the event horizon is less than $10^{-8}~M_\odot~\rm yr^{-1}$. For this work we used $\dot M \sim 10^{-8} M_\odot ~\rm yr^{-1}$ for the vicinity of the black hole. 

The ram pressure can be estimated as 
\begin{equation}
\label{eq:sigma_g}
\begin{aligned}
    \sigma_\g & \sim  \rho_\g v_{\rm rel}^2 \\
    & = 7500 \left({\dot M \over 10^{-8} {M_\odot ~\rm yr^{-1}}} \right) \left({R \over R_\bullet} \right)^{-5/2}  \rm ~Pa,
\end{aligned}
\end{equation}
where $\rho_{\rm g}$ is the gas density and $v_{\rm rel} \sim \sqrt{2 G M_\bullet / r}$ is the relative speed of an asteroid in a nearly parabolic orbit with respect to the gas disk. 

For rubble-pile bodies, the survival condition is $\sigma_{\rm g} < \sigma_{\rm c, rb} $, which gives
\begin{equation}
    R > 22~R_\bullet ~.
\end{equation}
This distance is slightly larger than the tidal disruption distance for rubble piles, indicating that rubble piles should become disrupted due to tides around $\sim$$22 R_\bullet$, within which small bodies can only exist in the form of monolithic objects. On the other hand, monolithic objects can easily survive from the gas pressure as their strength (e.g., >10~kPa) is much greater than $\sigma_{\rm g}$.

\section{Evaporation and erosion of fragments}
\label{sec3}

We  show that large objects can survive down to $22R_\bullet$, where they fragment into smaller bodies with sizes below 100~m as a result of gas pressure and tides. However, these small objects may undergo evaporation during a flyby due to frictional heating by the surrounding gas. In this section, based on the fireball model, we estimate the closest distances these small fragments can approach without being completely evaporated during a single passage, and we evaluate the resulting luminosity.

\subsection{Evaporation}

The aerodynamic friction experienced by a planetary object moving through the surrounding gas leads to surface heating and potential evaporation. We follow the approach of \citet{zubovas2012} to estimate the evaporation rate.
The mass-loss rate can be estimated by using the classical meteor ablation formula \citep{Bronshten1983}:
\begin{equation}
    \dot m \sim - \frac{C_{\rm H} \pi r^2 \rho_g v^3 }{ 2Q}.
\end{equation}
Here $C_{\rm H} \sim 1$ is a dimensionless factor that measures the fraction of the energy that is used for evaporation instead of thermal emission \citep{zubovas2012}.
The parameter $Q \sim 3 \times 10^{6} ~\rm J~kg^{-1}$ is the specific heat of ablation, which is the energy required to ablate a unit mass of the material \citep{Borovicka2007}. The radius shrinkage rate is
\begin{equation}
\label{eq:dot_r}
    \dot r = \frac{\dot m}{4\pi r^2 \rho} \sim 50 \left( \frac{R}{R_\bullet} \right)^{-3} \rm m~s^{-1}.
\end{equation}
Here we assume the density $\rho \sim 3000 ~\rm kg~m^{-3}$, which is the typical grain density of carbonaceous chondrite meteorites \citep{Flynn1999}. After the time spent by the object at a distance comparable to the pericenter radius, $t_{\rm fly} \simeq \pi \sqrt{R^3 / 2 G M_{\bullet}}$, the evaporation radius is $r_{\rm ev} = \dot r t_{\rm fly} \sim 100 (R/R_\bullet)^{-5/2}~$m.

The red curve in Fig.~\ref{fig:asteroid_survival} represents the depth of evaporation caused by frictional heating during a single pericenter passage, setting a lower limit on the size of surviving objects. This analysis suggests that the largest monolithic objects can reach  $\sim 8 R_\bullet$. This location is consistent with the flare observed at $\sim 7 R_\bullet$ \citep{Vincent2011, GC2023}.

These estimates assume that objects experience continuous gas friction throughout a flyby and that their velocity is not significantly reduced by gas drag. Below, we justify these assumptions. Objects with inclination $\sin i < H/R$ are embedded in the gas disk and continuously experience drag heating. Objects with $\sin i>H/R$ experience a heating time $\sim t_{\rm fly}  H/ \pi R $. However, considering the value of  $H/R \sim 1$ in the vicinity of the black hole, the objects that come from the stellar disk with $i < 45^\circ$ are expected to be embedded in the gas disk, and thus subject to sustained heating. If the aerodynamic drag slows the fragments on a timescale shorter than the evaporation timescale, the fragments will co-move with the ambient gas, effectively halting further evaporation. However, in Appendix~\ref{app1}, we show that the aerodynamic stopping time is much longer than the evaporation time.

\subsection{Discussions on luminosity}

The interaction between evaporated small-body fragments and the background accretion flow could generate strong emissions through different mechanisms, including high-energy electrons generated after sublimation and the thermal emission. Through high-energy electrons, \citet{zubovas2012} assumed that the power released by a fragment is a fraction of the rest-mass energy per unit time, i.e.,
\begin{equation}
    L_{\rm bulk} = { \xi mc^2  \over t_{\rm fly}}.
\end{equation}
Here $\xi \sim 0.1$ is the dimensionless fraction of the released rest-mass energy \citep{zubovas2012}. The pericenters of the fragments of the parent planetary scatter randomly around the disk and can be evaporated during one or a few passages to the black hole. The total bulk energy of the fragments is equal to that of the parent body. For a 10 km object, the luminosity could reach $3 \times 10^{35}~\rm erg~s^{-1}$.

Here we reinvestigate the scenario of thermal emission of fragments by adopting the fireball flare model. As the fragments spread out, each fragment heats up individually, increasing the total surface area exposed to the atmosphere (see, e.g.,  Fig.~5 in \citet{Hulfeld2021}). This could result in a sudden burst of light or flare as the fragments continue to ablate, releasing energy over a short time. A similar mechanism is responsible for the bright flashes observed during meteor fireballs in Earth’s atmosphere.

Simulations by \citet{Hulfeld2021} indicate that shortly after fragment formation, there is no mutual shielding among fragments, meaning that each fragment grain receives the full heat flux without obstruction. A commonly used approach to estimate the luminosity of each fragment is 
\begin{equation}
    I = - f \frac{v_{\rm rel}^2}{2} (4 \pi r^2 \rho \dot r ) ,
\end{equation}
where $f$ is the luminosity efficiency and $\dot r$ can be obtained from Eq.~(\ref{eq:dot_r}).

To determine the total luminosity produced by all fragments, an appropriate size distribution must be adopted. We implement the size distribution of meteor fragments, which has been studied extensively based on recovered meteoritic debris. Several distribution models have been considered in previous works, including a simple power law \citep{Frost1969}, a power law with an exponential cutoff \citep{Oddershede1998, Vinnikov2016, Gritsevich2017}, a third-degree polynomial \citep{Badyukov2013}, and bimodal distributions. Recent findings by \citet{Brykina2024a} indicate that the mass distribution of meteorite fragments and impact experiment debris is well described by a power law of the form $dN \propto m^{-\beta} dm$, which could translate to a size distribution of
\begin{equation}
    {\rm d}N \propto r^{-(3\beta -2)} dr.
\end{equation}
Applying the condition that the total mass of fragments equals the mass of the parent body, we have
\begin{equation}
    {\rm d}N = {\frac{(6 - 3\beta) r_0^3}{r_{\rm max}^{6-3\beta} - r_{\rm min}^{6-3\beta}}} r^{-(3\beta -2)} dr,
\end{equation}
where $r_0$ is the size of the parent body, and $r_{\rm min}$ and $r_{\rm max}$ are the minimum and maximum size of the fragments. The total luminosity can be calculated by summing up the energy released  by all the fragments,
\begin{equation}
    L_{\rm fireball} = \int I(r) {\rm d}N,
\end{equation}
which gives the relation between the parent body size, $r_0$, and the total luminosity, $L_{\rm fireball}$. In our calculation, we take $r_{\rm min} = 100 ~\rm \mu m$, as suggested by observations of meteors \citep{Borovicka2007}, and $r_{\rm max} =$ 100~m. The power index, $\beta$, is found to be $ \sim$ 1.45 – 1.7 for meteoritic fragments and $\sim$ 1.68 – 1.88 for impact experiment fragments \citep{Brykina2024a}. They also found there could be bias against small fragments in the collection, suggesting that the $\beta$ could become larger when more fragments are collected. Fragments produced via tidal disruption follow a different size distribution. 

Numerical simulations by \citet{Zhang20} demonstrate that the size distribution of tidal fragments cannot be described by a single power law. For large fragments with sizes exceeding 30\% of the parent body’s radius, the power-law index is as steep as $\beta \sim 3$, while for smaller fragments $\beta$ is $\sim 1.3$. This shallower distribution may result from the re-union of the fragments due to the self-gravity in the gas-free environment, which is not suitable when accounting for the disruption process induced by the gas flow around the black hole. Therefore, we implement the size distribution of meteor fragments instead of tidal fragments. We chose $\beta = 1.8$ for this study.

\begin{figure}
    \centering
    \includegraphics[width=\linewidth]{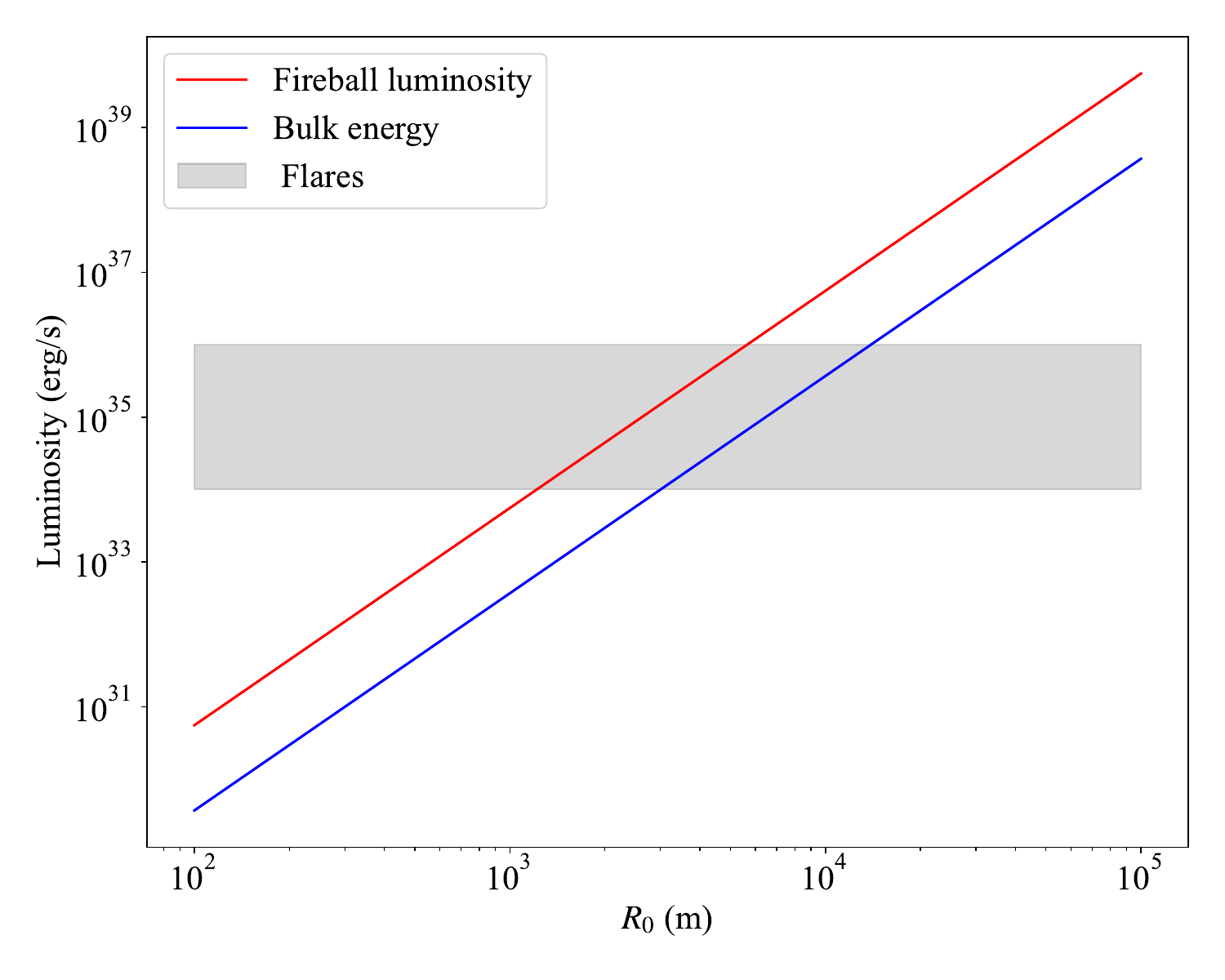}
    \caption{Luminosity produced by asteroids at $\sim$$8~R_\bullet$ based on the estimate of the bulk energy (red) and the fireball luminosity (blue). The luminosity of the observed flares is represented by the gray region.}
    \label{fig:luminosity}
\end{figure}

Figure~\ref{fig:luminosity} shows the luminosity generated by asteroids based on the two mechanisms discussed above. Our calculation indicates that the fireball luminosity produced by a parent body with radius $r = 10$~km could reach $5 \times 10^{36} \rm ~erg/s$ at $~ 8 R_\bullet$, typically one order of magnitude higher than the bulk energy luminosity. As illustrated in Fig.~\ref{fig:luminosity}, planetary bodies with radii from a few kilometers to ten kilometers are sufficient to reproduce the observed flare luminosities. In contrast to \citet{zubovas2012}, we show that the thermal emission from fragments could reproduce the observed luminosity using the fireball model adopted in meteor luminosity studies. However, it is likely that multiple emission mechanisms contribute to the observed flares. Our estimated size requirement is slightly smaller than their 10~km estimate.

While the exact emission mechanism remains uncertain, we can compare the statistical properties of observed flares with those expected from asteroid-induced emissions. The flare frequency--luminosity relation is described by ${\rm d}N_{\rm f} \propto L^{-\alpha_{\rm L}}$. Statistical studies suggest $\alpha_{\rm L}$ to be 1.9 \citep{neilsen2013},  1.8 \citep{Wang2015}, 1.65 \citep{Li2015}, and 1.73 \citep{Yuan2018}. For asteroid-induced flares, luminosity is proportional to mass. Under a collisional cascade, the asteroid mass distribution follows a power law with index 1.83, implying $\alpha_{\rm L} = {1.83}$, consistent with the observational range.

The duration distribution of flares does not follow a clear power law \citep{Yuan2018}, but may be approximated by $t_{\rm f} \propto L^{-\alpha_{t}}$ with $\alpha_{\rm t} < 0.55$ \citep{Li2015}. In the asteroid emission scenario, the flare duration corresponds to the evaporation time, $t_{\rm evap} = r / \dot r $. Since $\dot r$ is independent of $r$ (Eq.~\ref{eq:dot_r}), we find $t_{\rm f} \propto r \propto m^{-1/3} \propto L^{-1/3}$, consistent with constraint of $\alpha_{\rm t} < 0.55$. 

\section{Delivery of planetary bodies to Sgr A$^\star$}
\label{sec4}

Sgr A$^\star$ is surrounded by a nuclear cluster of $N_\star \sim M_\bullet/M_\star$ 
old stars \citep{schodel2007, trippe2008} with solar or subsolar masses $M_\star$.
Under the dominance of the SMBH's gravity, these stars undergo scalar resonant relaxation 
\citep{rauch1996, hopman2006, yu2007}, which redistributes angular momentum among 
themselves. On a secular (scalar relaxation) timescale $\tau_{\rm secular} 
\sim P_\star M_\bullet/M_\star (\sim 10^9$ yr at 0.1pc), 
a small fraction of stars attain a sufficiently 
small pericenter distance $R_{\rm peri} (\lesssim R_{\rm t, grav}$, Eq. \ref{eq:rgrav}) to undergo tidal disruption \citep{rees1988}.  
Under the assumption that the cluster's dynamical structure evolved concurrently 
with the growth of the SMBH's mass, $M_\bullet$, the stellar tidal disruption rate 
around Sgr A$^\star$ is expected to be $\sim 10^{-5}$ yr$^{-1}$ \citep{alexander2005}.
Since close-in super-Earths are common around solar-type stars \citep{kunimoto2020, bryson2021},
they may accompany their host stars destined to the SMBH proximity {{at a rate of 
$\sim 10^{-11} M_\odot$ yr$^{-1}$, corresponding to a rest-mass energy flux of a few $\times 10^{35} \rm erg~s^{-1}$, sufficient to account for the observed flare energetics.}}

Host stars cannot retain planets or small bodies with a semimajor axis $a_{\rm sb} 
\gtrsim R_{\rm peri} (M_\star/M_\bullet)^{1/3}$.  For the main belt-like 
asteroids and Kuiper belt-like objects, the detachment distance is $ R_{\rm peri} 
\sim 400~\rm AU > > R_{\rm t, grav}$ and $ \sim 7000~\rm AU$, respectively.
For super-Earths with periods that range from weeks to months, the detachment distance is
$R_{\rm peri} \sim$ a few AU, well outside $R_{\rm t, grav}$.
After their detachment from their host stars, resonant relaxation leads some
freely floating planetary bodies, including small bodies and gas giants, to $R_{\rm t, grav}$ for tidal disruption
of their gaseous envelopes and to $R_{\rm t, str}$ (Eq. \ref{eq:rstr})
for tidal downsizing of their solid cores.
\citet{zubovas2012} reproduced the delivery rate inferred from the observed 
outbursts under the assumption that all the cluster stars carry 
a population of $N_{\rm sb}$ outlying (beyond the ice giant planets) small 
bodies with a total mass $M_{\rm total} \sim 10^{-5} M_\odot$.  Their
time-average detachment rate is $\sim N_\star N_{\rm sb} / \tau_{\rm 
secular}$.  The assumed magnitude of $M_{\rm total}$ is comparable to that
of super-Earths and long-period Neptune-mass planets \citep{zang2025}.  It is 
a fraction that of the primordial Kuiper belt required ($\sim 6 \times 
10^{-5} M_\odot$) to establish the current kinematic structure of the 
outer Solar System \citep{Gomes2005, Nesvorny2018}. Comparable masses have 
been inferred for debris disks around other main sequence stars, 
including $\beta$ Pictoris.  It is also within the range needed 
to continually pollute the atmospheres of some white dwarfs \citep{zhang2021c}. 

Although concurrent mass segregation is associated with dynamical friction (also leading to a gradual depletion of freely floating small bodies and planets from the nuclear 
cluster), it proceeds on a characteristic timescale of $\tau_{\rm loss} = 10^9~$yr,
which is $\sim \tau_{\rm secular}$. In a continuous detachment-loss equilibrium, 
the population of small bodies in the nuclear cluster remains relatively 
constant over time.

Here we consider another 
potential supply that increases the tidal disruption rates of stars and planetary bodies.
Sgr A$^\star$ is surrounded by a group of young S stars \citep{Ghez2003a, gillessen2017} 
with highly eccentric and isotropic orbits at $5 \times 10^{-3}-0.04$pc, a 
clockwise rotating disk of O/WR stars (CWSs) with modest orbital eccentricities, and 
some off-the-disk B stars (ODSs) with high eccentricity and inclination (relative to the 
clockwise-rotating disk) extends from approximately 0.04~pc to 0.5~pc 
\citep{Levin2003, paumard2006, lu2009, yelda2014, ali2020, vonfellenberg2022}. 
These young massive stars ($\sim$ 200 in number)  either recently formed
through gravitational instability\citep{Goodman2003}   or rejuvenated after being captured \citep{artymowicz1993, davies2020} by a gaseous disk around the SMBH. These 
young stars are likely to be surrounded by their own protoplanetary disks, some of which may 
have persisted for several megayears and manifested as  G-clouds \citep{Gillessen2019, 
owen2023}.  The planet formation probability in these protoplanetary disks may 
be increased \citep{idalin2004b} by 
the supersolar metallicity commonly found in typical active galactic nucleus (AGN) disks \citep{huang2023}.  
Moreover, entities with masses comparable to planets, also known as blanets, 
may also form {in situ} and concurrently with emerging stars,
directly within the disk around the SMBH \citep{Wada2019, Wada2021}.  

Although the natal disk of S stars, CWSs, and ODEs no longer exists in the 
proximity of Sgr A$^\star$ today, its surface density and 
temperature distribution may be inferred from AGN disks around 
other SMBHs with similar $M_\bullet$. These quantities have been 
extrapolated from the reverberation mapping and radiation 
transfer models of broad emission lines and spectral energy 
distribution of typical AGN disks.  The interaction of
embedded and surrounding stars with their natal disk increases 
their in-spiral and tidal disruption rates \citep{macleod2020, wang2024}.
During the depletion of their natal disk, the stars' eccentricity $e_\star$ is 
further excited, together with a decline in their $R_{\rm peri}$ by
either the von Zeipel-Lidov-Kozai effect \citep{valtonen2006} 
or sweeping secular resonance \citep{zheng+2020, zheng2021} of any 
intermediate-mass companion to the SMBH, such as IRS 13E \citep{Krabbe1995}.

In comparison with the old nuclear cluster stars, these young stars 
have higher masses ($M_\star \sim 15 M_\odot$). With their periods of decades to centuries, S stars' $\tau_{\rm secular} \sim 10^7$ yrs. 
Many ($N_{\rm S}>20$) 
S stars also have $R_{\rm peri} \lesssim 10^{-2}$ pc \citep{burkert2024}.
Around them, even planets with semimajor axes less than that of the ice giant 
planets in the Solar System and some known cold Jupiters around other stars
in the solar neighborhood can also become detached from their
hosts.  These escapers form a super-Oort cloud composed of a population of 
planets and small bodies detached from their original host stars
\citep{Nayakshin2012}.  These free-floating planetary 
bodies provide a rich reservoir for subsequent 
planetary tidal disruption of their gaseous envelope 
and downsizing of their solid cores when they venture to 
the SMBH's $R_{\rm t, grav}$ and $R_{\rm t, str}$.
Under the assumption that short-period super-Earths and long-period 
Neptunes are common \citep{kunimoto2020, bryson2021, zang2025}, 
we estimate the total mass of the super-Oort cloud to be $M_{\rm SO} 
\gtrsim N_{\rm S} 
M_\oplus \sim 10^{28-29}$ g. From these estimates, we infer a time-averaged 
delivery rate $\sim M_{\rm SO}/\tau_{\rm secular} \sim 10^{13-15}$g 
s$^{-1}$, which exceeds that needed to account for the flares.

\section{{Emission signatures from ionized asteroidal materials}}

\begin{figure}
    \centering
    \includegraphics[width=\linewidth]{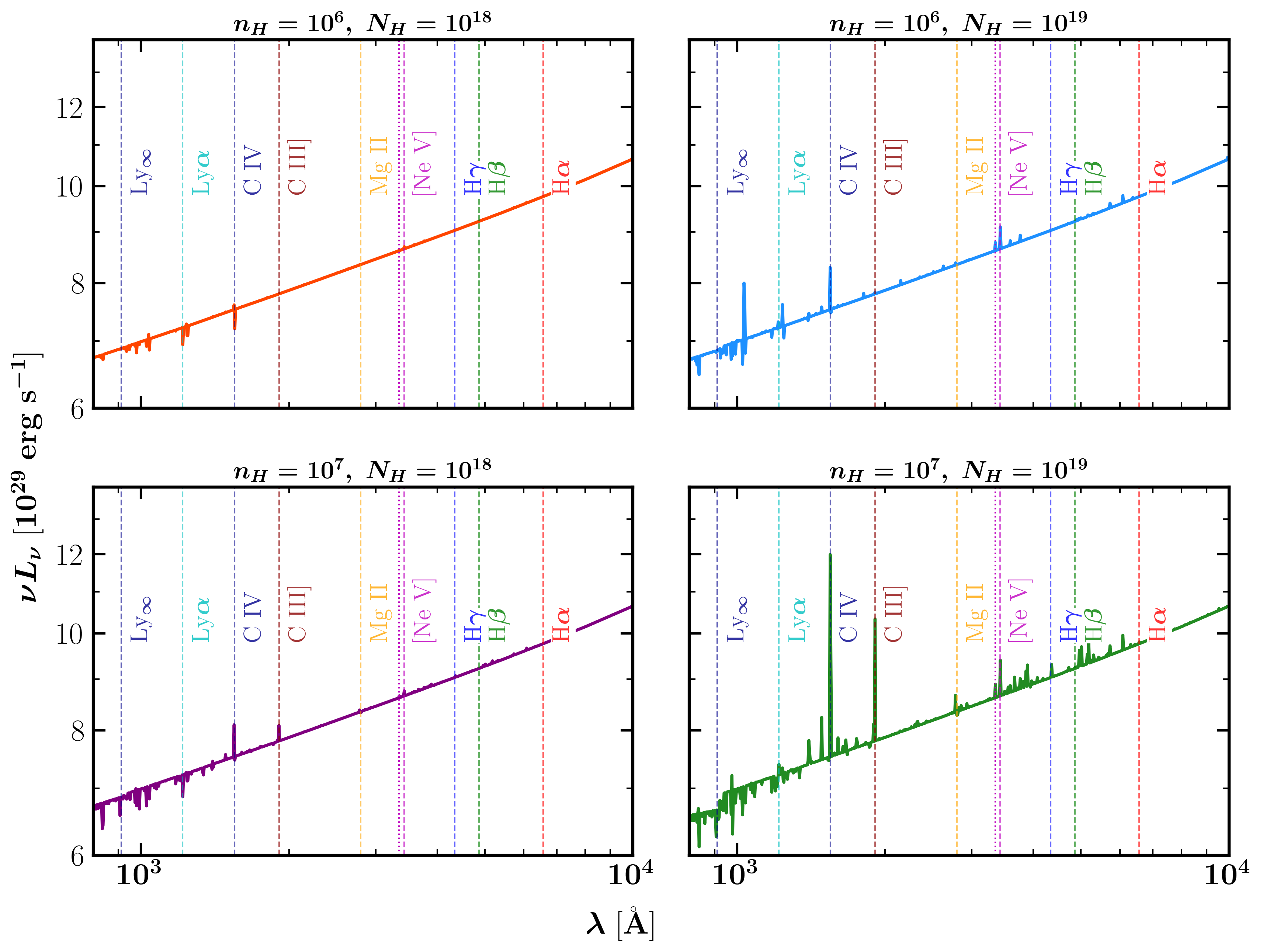}
    \caption{Predicted UV–optical emission spectra of photoionized asteroid material irradiated by ionizing continuum with $\phi = 2\times10^{15}~\mathrm{photons~s^{-1}~cm^{-2}}$. The four panels correspond to different combinations of gas density, $n_{\mathrm{H}} = 10^{6}$ (upper two panels) or $10^{7}~\mathrm{cm^{-3}}$ (bottom) and total hydrogen column density, $N_{\mathrm{H}} = 10^{18}$ (left) and $10^{19}~\mathrm{cm^{-2}}$ (right). Prominent emission lines are indicated by vertical dashed lines. We adopt a metallicity of $Z = 100\, Z_{\odot}$ and a covering factor of 5\%. These spectra illustrate how the emission-line strengths vary with gas density and column depth in the evaporating asteroidal clouds.}
    \label{fig:cloudy_spec}
\end{figure}

To investigate the observable spectral signatures of the post-flare under the asteroid disruption hypothesis, we computed photoionization models of the evaporated asteroidal gas using \textsc{Cloudy}.  We assumed that the fragments are exposed to the ionizing continuum associated with the observed flare SED of the Galactic center,  scaled such that the incident ionizing flux at the cloud surface is $\phi = 2\times 10^{15}$ photons s$^{-1}$ cm$^{-2}$ (e.g., Fig.1 in \citealt{owen2023}).  We assume that the composition of the gas reflects the supersolar metallicity expected for asteroid material, with ${Z = 100\,Z_{\odot}}$. The mean density of the evaporated material can be approximated as $\rho_{\rm ast} = m_{\rm ast} / v_{\rm rel} c_{\rm s}^2 t_{\rm evp}^3$. For a representative case with $m_{\rm ast} = 10^{19}~$g, $v_{\rm rel} = 0.3c$ at $10~R_\bullet$, $c_s \leq 10^8~\rm cm~s^{-1}$, $t_{\rm evp} = 30~$min, we obtain $\rho_{\rm ast} \sim 1.5 \times 10^{-17}~\rm g~cm^{-3}$, which is equivalent to $n_{\rm H} \leq 10^7 ~\rm cm^{-3}$. Four representative models are shown in Figure~\ref{fig:cloudy_spec}, corresponding to hydrogen densities $n_{\mathrm{H}} = 10^{6}$ and $10^{7}$ cm$^{-3}$ and total hydrogen column densities $N_{\mathrm{H}} = 10^{16}$ and $10^{17}$ cm$^{-2}$.  A covering factor of $C = 0.05$ was adopted to account for the small projected area of the fragment swarm. In these models, the dominant emission features include \ion{C}{iv} $\lambda1550$ and \ion{C}{iii} $\lambda1909$. For the higher-density and higher-column models, \ion{Fe}{ii} blend and \ion{[Ne}{V]} also become prominent. 

As shown in Figure~\ref{fig:cloudy_spec}, the emission-line luminosities systematically increase with gas density. We estimate  the Strömgren radius
($R_{\mathrm{S}} \propto \phi / (n_{\mathrm{H}}^{2}\alpha_{\mathrm{B}})$), $R_{\mathrm{S}} = 2.3\times10^{16}$ and $2.3\times10^{14}$ cm for 
$n_{\mathrm{H}} = 10^{6}$ and $10^{7}$ cm$^{-3}$, respectively. 
The corresponding ionized columns are $N_{\mathrm{S}} \simeq 2\times10^{22}$ and 
$2\times10^{21}$ cm$^{-2}$, both far exceeding the adopted model columns implied by the size and density of the asteroidal material. 
This implies that the clouds are fully ionized in all cases, and therefore recombination lines are generally weak in the predicted spectra. 
Nevertheless, the higher-density models radiate more efficiently because the emissivity of collisionally excited lines scales approximately as $n_{\mathrm{H}}^{2}$, while the stronger cooling at high density lowers the equilibrium temperature ($T_{\mathrm{e}}\sim10^{4}$~K) and increases the efficiency of metal-line excitation. 
Consequently, intermediate ions such as C$^{2+}$ and C$^{3+}$ become more abundant, producing prominent \ion{C}{iii]} and \ion{C}{iv} features. 
The [\ion{Ne}{v}]~$\lambda3426$ line also becomes stronger with both higher density and larger column depth, since a larger column simply provides more emitting material within the fully ionized region.
Together, these effects indicate that the enhanced \ion{C}{iv}, \ion{C}{iii]}, and [\ion{Ne}{v}] emission in the high-density model.

If we focus on the model that produces the strongest emission lines ($n_{\rm H} = 10^7~\rm cm^{-3}$ and $N_{\rm H} = 10^{19}~\rm cm^{-2}$), the most prominent feature in the UV spectrum is the \ion{C}{iv} line. The luminosities of individual lines are obtained by multiplying the \textsc{Cloudy} surface flux by the illuminated surface area and the covering factor, yielding $L_{\ion{C}{iv}} = 5.04\times10^{27}~\rm erg~s^{-1}$. This corresponds to an observed flux of $6.3\times10^{-19}~\rm erg~s^{-1}~cm^{-2}$. For comparison, the 5$\sigma$ line detection limit of HST/Cosmic Origins Spectrograph \citep[COS][]{Green2012} using the low-resolution grating G140L, which covers the \ion{C}{iv} wavelength range, is approximately $10^{-15}~\rm erg~s^{-1}~cm^{-2}$ for a typical total integration time of $\sim 1800~$s \citep[see, e.g.,][]{Anderson2016}. Therefore, even the brightest predicted UV line would remain well below the current detection threshold.

In summary, the metal line emission has a characteristic luminosity of about $10^{29}$ to $10^{30}~\rm erg~s^{-1}$, roughly five orders of magnitude lower than the typical emission from Sgr A$^\star$ ($10^{35}$ to $10^{36}~\rm erg~s^{-1}$). Consequently, any emission lines produced by vaporized asteroid material near the Galactic center would be far too faint to detect with the present-day instruments.

\section{Summary and discussions}
\label{sec5}

Sgr A$^\star$, the supermassive black hole at the center of the Milky Way, exhibits frequent short-duration flares across multiple wavelengths. The origin of these flares remains an open question. In this work, we revisited the mechanism involving the tidal disruption and subsequent evaporation of small planetary bodies. We refined the previous models by incorporating the material strengths constrained from recent space missions and dynamical studies.

We first analyzed the fragmentation of planetary bodies due to tidal forces and gas pressure. Large rubble-pile asteroids (>100~m) with cohesion $\sim 1~$Pa can survive down to $\sim 22 R_\bullet$, within which they break into monolithic fragments (<100~m) of much higher material strength. These monolithic objects, possessing higher material strength, can withstand further tidal disruption, but may still be subject to evaporation via aerodynamic heating during their close passage around the black hole. We estimated the surface temperature and mass-loss rate of these fragments, demonstrating that the strongest monolithic bodies can survive as close as $\sim 8 R_\bullet$ over one passage. This region is more consistent with observed locations ($\sim 7 R_\bullet$) of flaring events than $\sim 20 R_\bullet$ predicted by the previous study, which  neglected material strength \citep{zubovas2012}.

We examined the luminosity generated by small bodies and considered two emission mechanisms: (1) the bulk energy emission mechanism from high-energy electrons produced during sublimation, suggested by \citet{zubovas2012}, and (2) thermal emission modeled as fireball flares akin to meteor events in Earth's atmosphere. In the fireball model, each fragment heats independently, releasing energy as surface area increases. Using a physically motivated fragment size distribution drawn from meteorite and impact studies, we integrated individual fragment luminosities to estimate the total emission. Our results show that a  parent body of a few kilometers in size can produce fireball luminosities comparable to the flare luminosity. 

Statistical comparisons with observed flares support the asteroid-induced flare hypothesis. The flare frequency--luminosity relation, expected to follow ${\rm d}N \propto L^{-\alpha_{\rm L}}$ with $\alpha_{\rm L} \sim 1.83$, agrees with the observed values \citep[$\alpha_{\rm L} \sim 1.65-1.9$; e.g.,][]{neilsen2013, Wang2015, Li2015, Yuan2018}. Additionally, the flare duration-luminosity scaling ($t_{\rm f} \propto L^{-\alpha_{\rm t}}$) with $\alpha_{\rm t} \sim 1/3$, derived from fragment evaporation times aligns with observational constraints $\alpha_{\rm t} <0.55$ \citep{Li2015}. {{We found that the metal emission in the post-flare phase has a characteristic energy flux of $\nu L_\nu\sim 10^{29}-10^{30}  \rm ~erg~s^{-1}$, five orders of magnitude lower than the emission of Sgr A$^\star$ (i.e., $10^{35}-10^{36} \rm ~erg~s^{-1}$). Consequently, any emission lines produced by the vaporized asteroid material would be far too faint to be detected.}}

Finally, we consider the delivery of these kilometer-scale small bodies to the Roche limit of Sgr A$^\star$. These objects, originally from planetary systems around S stars, CWSs, ODSs, and nuclear cluster stars,  are transported along with their host stars to  Sgr A$^\star$'s proximity thanks to  Sgr A$^\star$'s  dominant gravity and stellar resonant relaxation. 
After their detachment from their host stars, their orbits continue to evolve through resonant 
relaxation.  Free-floaters lose their gaseous envelope when their pericenter distance 
is reduce to $\lesssim r_{\rm t, grav}$.  Eventually, their residual solid cores undergo 
tidal downsizing when their $R_{\rm peri}$ is reduced to $r_{\rm t, str}$.
Assuming a total small-body mass comparable to that of the primordial Kuiper belt in the Solar 
System and the population of exoplanets, the delivery rate to the vicinity of the black hole could reach several per day, consistent with the frequency of observed flares.

\begin{acknowledgements}
We would like to thank Eiichiro Kokubo for the useful discussion. W.H. Zhou gratefully acknowledges the support and supervision received at the Observatoire de la Côte d’Azur, where the majority of this work was carried out. W.H.Zhou is currently affiliated with the University of Tokyo.
\end{acknowledgements}

\bibliographystyle{aa} 
\bibliography{references} 

\appendix

\section{Aerodynamic stopping timescale}
\label{app1}

The aerodynamic stopping timescale for a fragment due to gas drag is 
\[
t_{\rm s} = 
\begin{cases}
  {\rho r}/{\rho_{\rm g} v_{\rm rel}}  & \text{if $r < 9\lambda/4$} \\
  4 \rho r^2 / 9\rho_{\rm g} v_{\rm rel} \lambda & \text{if $r > 9\lambda/4$, Re$<1$}
\end{cases}
\]
when their relative speed is faster than the sound speed \citep{Chiang2010}. Here Re$ = r v_{\rm rel}/\lambda c_{\rm g}$ is the Reynolds number. The mean free path of the gas $\lambda$ is roughly
\begin{equation}
    \lambda = \frac{m_{\rm H}}{\rho_g \sigma_{\rm H}} \sim 6.8 \times 10^7 \left({R \over R_\bullet} \right)^{3/2}   ~ \rm m,
\end{equation}
where $m_{\rm H} = 1.67 \times 10^{-24}~$g is the mass of the hydrogen atom and  $\sigma_{\rm H} \sim 3 \times 10^{-15}~\rm cm^2$ is the effective collisional cross section. The Epstein regime applies to planetary bodies, for which the stopping time can be estimated as 
\begin{equation}
    t_{\rm s} \sim 10^9 t_{\rm fly} \left({r \over 1~{\rm km}} \right) \left({R \over R_\bullet} \right)^{1/2}.
\end{equation}
It turns out that the stopping time is so long that fragments of all sizes evaporate before they are coupled with the ambient gas. 

\end{document}